\documentclass[aps,pra,preprint,superscriptaddress]{revtex4-2}
\usepackage{braket}
\usepackage{graphicx}
\usepackage{amsmath}
\usepackage{amssymb}
\usepackage{amsfonts}
\usepackage{url}
\usepackage{xcolor}
\usepackage[
    colorlinks = true,
    urlcolor   = blue,     
    linkcolor  = red,      
    citecolor  = green     
]{hyperref}

\begin{document}

%\preprint{}

\title{Enhancing Quantum Machine Learning with Anyons}
\author{Da Zhang}
    \affiliation{Center for Quantum Technology Research and Key Laboratory of Advanced Optoelectronic Quantum Architecture and
Measurements (MOE), \\ School of Physics, Beijing Institute of Technology, Beijing 100081, China}
    
\author{Wen-Qiang Liu}
\affiliation{Department of Mathematics and Physics, Shijiazhuang Tiedao University, Shijiazhuang 050043, China}

\author{Zhaohui Wei}
\affiliation{Yau Mathematical Sciences Center, Tsinghua University, Beijing 100084, China}
\affiliation{Yanqi Lake Beijing Institute of Mathematical Sciences and Applications, Beijing 100407, China}
  
\author{Zhang-Qi Yin}
    \email{zqyin@bit.edu.cn}
    \affiliation{Center for Quantum Technology Research and Key Laboratory of Advanced Optoelectronic Quantum Architecture and
Measurements (MOE), \\ School of Physics, Beijing Institute of Technology, Beijing 100081, China}

%\date{\today}

\begin{abstract}
The power of quantum computing and quantum machine learning relies on harnessing uniquely quantum phenomena as computational resources. While superposition, coherence and entanglement have been central to this effort, the role of particle exchange statistics remains largely unexplored. Here, we introduce a quantum kernel framework that unifies bosonic, fermionic, and anyonic (fractional) exchange statistics within a single learning paradigm. We study this family of kernels from three perspectives. At the representation level, Haar-averaged effective-dimension analysis shows that fractional exchange phases access feature-space directions inaccessible to the purely symmetric or antisymmetric limits. At the level of kernel geometry, the corresponding Gram matrices show greater separation from the distinguishable-particle baseline and reduced label-dependent model complexity. Finally, on learning benchmarks, anyonic kernels consistently outperform their bosonic and fermionic counterparts, with stronger target alignment and more favorable class geometry. Together, these findings show that exchange statistics reshape the structure and geometry of quantum feature space, leading to enhanced learning performance. Our work identifies particle exchange statistics as an overlooked computational ingredient for quantum machine learning and provides the first systematic comparison of quantum learning models across exchange phases.

\end{abstract}

\maketitle

\section{Introduction}

Quantum computation aims to harness fundamental quantum-mechanical principles into computational power. In recent years, this idea has evolved from a theoretical possibility into experimental demonstrations, including random-circuit sampling with programmable superconducting processors, photonic boson sampling, and large-scale simulations of many-body dynamics~\cite{2019supermacy,2025fansuper,gbs2017,lugbssupermacy2020,alam2025fermionic,alam2025fermihubbard}.
These experiments exploit phenomena with no direct classical counterpart, such as coherent superposition, many-body entanglement and multiphoton interference among identical particles\cite{superposition,RMPcoherence,entanglerole,zhao2025entanglement,interference2014}.
These developments suggest a broader principle: the intrinsic physical properties of quantum systems can be converted into computational resources when embedded in an appropriate information-processing architecture.

Quantum machine learning provides a natural setting in which to ask whether physical properties of quantum systems can become useful for data analysis~\cite{qml17,cerezo2022qml}. One established route is to impose symmetry directly on the learning model. In classical machine learning, this principle underlies permutation-invariant and group-equivariant architectures~\cite{deepsets,cohen2016gcnn}, while covariant quantum kernels have shown how group structure in data can be encoded into quantum feature spaces~\cite{glick2024covariantkernel,Agliardi2025cov}. These approaches predominantly rely on top-down mathematical constraints, where symmetries must be artificially engineered into the model architecture or the embedding map. This contrast reveals a fundamentally distinct and deeply physical paradigm where the intrinsic properties of the quantum particles themselves naturally supply the required symmetry structure.

Particle exchange statistics~\cite{pauli,leinaas1977identical} is a natural candidate for such an intrinsic symmetry structure, yet it remains much less explored as an elementary property of quantum systems in quantum learning.
Exchange statistics specifies how a many-particle wave function responds when identical particles are exchanged.
Bosons and fermions correspond to the two familiar endpoints, with symmetric and antisymmetric exchange responses.
More general exchange phases are possible for Abelian anyons in two-dimensional topological matter, where an exchange can imprint a fractional statistical phase~\cite{wilczekanyon,wilczekstatistic82,hallanyon97}.
Recent interferometric experiments in fractional quantum Hall devices have made this exchange phase directly observable through anyonic interference and braiding signatures~\cite{lawmzi06,nonabelianmzi,2023anyonmzi,ghosh2025mzi}.
This progress motivates a simple question for quantum learning: Can exchange statistics be turned from a classification of particles into a design principle for learning machines?

Here we address this question using a photonic simulation route, in which entangled photons and linear optics emulate anyon-like exchange statistics~\cite{entanglewalk06,entanglewalk07,matthews2013entangle,prlgeneralwalk}.
On this basis, we construct a quantum kernel family that continuously interpolates between bosonic, fermionic and fractional-statistics exchange responses\cite{kernelfeaturesapce,QuantumEnhancedkernel,supervisedkernel,expkerneldu,speedupkernel}. The resulting framework provides a controlled setting for studying how an intrinsically quantum symmetry structure reshapes the geometry of a feature map, the complexity of the induced learning model and its predictive performance~\cite{Huangpower}.
Scanning the exchange phase across this kernel family allows us to track how particle statistics reshapes the data-dependent kernel geometry and the resulting learning behaviour.
We find that intermediate fractional phases produce advantageous kernel geometries: the effective dimension is enhanced, the separation from a classical, distinguishable-particle kernel is increased, and the label-dependent model complexity is reduced.
To demonstrate practical relevance, we benchmark these kernels on standard classification datasets, MNIST and Fashion-MNIST~\cite{lecun-mnist,Fasion-MNIST}.
Across these tasks, kernels incorporating anyonic exchange consistently outperform both bosonic and fermionic limits.
These results identify exchange statistics as a fundamental ingredient that can enhance quantum machine learning, suggesting that the structure imposed by particle exchange may play a wider role in quantum information processing and simulation.

\section{Fractional-statistics Quantum Kernel}

\begin{figure*}[htp]
    \centering
    \includegraphics[width=0.98\linewidth]{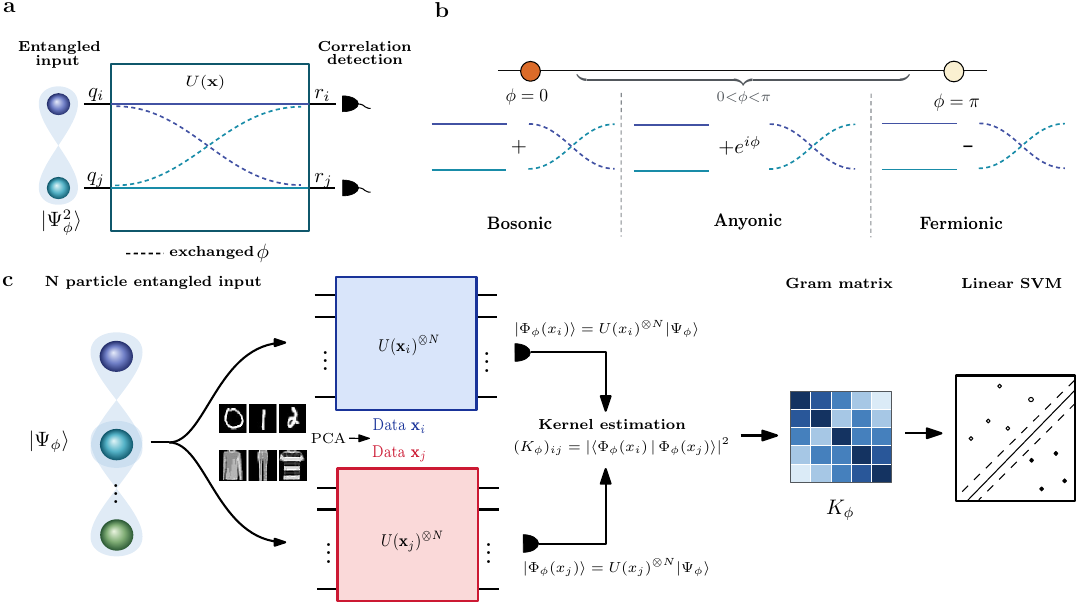}
    \caption{
\textbf{a}, Two-particle implementation of the encoded exchange phase. 
The entangled input contains direct and exchange components, which are mapped by $U(\mathbf{x})$ and read out by label-unresolved correlation detection, giving the two-path response in equation~\eqref{eq:two_particle_probability}.
\textbf{b}, Exchange responses selected by the encoded phase $\phi$. 
The cases $\phi=0$, $0<\phi<\pi$, and $\phi=\pi$ correspond to bosonic, anyonic, and fermionic responses, respectively.
\textbf{c}, $N$-particle fractional quantum kernel workflow. 
PCA-preprocessed data are encoded into programmable linear-optical unitaries $U(\mathbf{x})$. 
Acting on the permutation-entangled input $|\Psi_{\phi}\rangle$, these unitaries generate fractional-statistics feature states. 
Their pairwise overlaps define the kernel Gram matrix $K_\phi$ used by a classical SVM with a precomputed quantum kernel.
    }
    \label{fig:frac_kernel}
\end{figure*}

Exchange symmetry is a defining characteristic of quantum systems, providing the underlying mechanism that reshapes many-particle interference. In nature, this exchange phase is restricted to $0$ and $\pi$ for bosons and fermions. However, more general exchange statistics can be realized if the amplitude of the many-body system under particle exchange carries a tunable phase. Although naturally associated with anyonic excitations, these fractional statistics can be effectively engineered and simulated using photonic platforms~\cite{prlgeneralwalk,matthews2013entangle,Qianggeneralwalk}.

Figure~\ref{fig:frac_kernel}a illustrates the two-particle case.
The measured correlations arise from two indistinguishable alternatives: the direct assignment of inputs to outputs, and the exchanged assignment.
For two input modes $q_i$ and $q_j$, the initial state coherently superposes these alternatives with a tunable exchange phase:
\begin{equation}
|\Psi_{\phi}^{(2)}\rangle = \frac{1}{\sqrt{2}}\left(\hat a_{q_i}^{\dagger(1)}\hat a_{q_j}^{\dagger(2)} + e^{-i\phi}\hat a_{q_j}^{\dagger(1)}\hat a_{q_i}^{\dagger(2)}\right)|0\rangle .
\label{eq:two_particle_input}
\end{equation}
Here, the superscripts label the internal degrees of freedom of the two photons.
After both components evolve through the same data-dependent linear-optical unitary $U(\mathbf{x})$, the correlation measurement detects only the output modes, leaving the internal labels unresolved.
The two alternatives therefore interfere coherently.

The corresponding two-mode correlation probability is
\begin{equation}
P_{r_i r_j}^{\phi}(\mathbf{x})
= \frac{1}{2}
\Bigl|
\underbrace{
U_{r_i q_i}(\mathbf{x})U_{r_j q_j}(\mathbf{x})
}_{\text{direct}}
+
\underbrace{
e^{-i\phi}U_{r_i q_j}(\mathbf{x})U_{r_j q_i}(\mathbf{x})
}_{\text{exchanged}}
\Bigr|^2 .
\label{eq:two_particle_probability}
\end{equation}
Intermediate phases $0<\phi<\pi$ produce anyon-like exchange statistics.
This corresponds to the two-particle Abelian anyon case, in which a single generator of $B_2$ contributes a phase $e^{iv\theta}$ to the exchanged amplitude~\cite{artin1947braid,wilczk2024fractional}.
Here, $\theta$ is the statistical exchange angle acquired in a single elementary exchange, and $v$ counts the number of such exchanges.
The photonic $S_2$ construction implements the identity and transposition alternatives, weighting the latter by $e^{-i\phi}$.
In this way, Abelian-anyon–like interference is realized without physical braiding, with the phase encoded in the entangled input and revealed through label-unresolved correlation detection.

The two-particle construction above shows how an exchange phase can be programmed and read out through interference.  
We now use it to build a quantum kernel in which fractional exchange statistics is embedded directly into the feature map.  
As illustrated in Fig.~\ref{fig:frac_kernel}c, we generalize the direct--exchange superposition to an $N$-particle state defined over particle-to-mode assignments.  
For an input spatial configuration $\mathbf q = (q_1, \ldots, q_N)$, we prepare
\begin{equation}
|\Psi_{\phi}\rangle =
\frac{1}{\sqrt{N!}}
\sum_{\sigma \in S_N}
e^{-i \phi \pi(\sigma)}
\prod_{k=1}^{N} \hat a^{\dagger(k)}_{q_{\sigma(k)}} |0\rangle .
\label{eq:frac_input_state}
\end{equation}
Here, $S_N$ is the permutation group, $\pi(\sigma)$ is the inversion number of $\sigma$, and $\hat a^{\dagger(k)}_{q}$ creates a photon with internal label $k$ in spatial mode $q$.  
The input modes are assumed distinct, $q_i \neq q_j$ for $i \neq j$.  
For $N=2$, equation~\eqref{eq:frac_input_state} reduces to equation~\eqref{eq:two_particle_input}.  
For larger $N$, the phase factor $e^{-i \phi \pi(\sigma)}$ assigns a controllable phase to each permutation according to the number of adjacent exchanges needed to generate it.  
Because the measurement does not resolve the internal labels, these permutation components interfere coherently in the output.

To construct the quantum kernel, classical data points $\mathbf{x}$ are first compressed via principal component analysis (PCA) and subsequently encoded into a programmable linear-optical unitary transformation $U(\mathbf{x})$. Applying this unitary to an $N$-particle vacuum state yields the quantum feature state
\begin{equation}
|\Phi_{\phi}(\mathbf{x})\rangle =
\frac{1}{\sqrt{N!}}
\sum_{\sigma\in S_N}
 e^{-i\phi\pi(\sigma)}
\prod_{k=1}^{N}
\left(
\sum_{r_k}U_{r_k,q_{\sigma(k)}}(\mathbf{x})
\hat a^{\dagger(k)}_{r_k}
\right)|0\rangle .
\label{eq:expanded_feature_state}
\end{equation}
The overlap between two such feature states evaluates to
\begin{equation}
\langle \Phi_{\phi}(\mathbf{x}_i)|\Phi_{\phi}(\mathbf{x}_j)\rangle
=
\frac{1}{N!}
\sum_{\sigma,\tau\in S_N}
e^{i\phi[\pi(\sigma)-\pi(\tau)]}
\prod_{k=1}^{N}
M_{q_{\sigma(k)},q_{\tau(k)}}(\mathbf{x}_i,\mathbf{x}_j),
\label{eq:double_perm_overlap}
\end{equation}
where $M(\mathbf{x}_i,\mathbf{x}_j)=U^\dagger(\mathbf{x}_i)U(\mathbf{x}_j)$ represents the relative single-particle unitary. By introducing the relative permutation $\alpha=\tau\circ\sigma^{-1}$, the double sum reduces to a single-sum interference representation. This rigorously isolates the coherent contributions from distinct permutation structures, yielding
\begin{equation}
\langle \Phi_{\phi}(\mathbf{x}_i) |
\Phi_{\phi}(\mathbf{x}_j) \rangle
=
\frac{1}{N!}
\sum_{\alpha \in S_N}
C(\alpha,\phi)
\prod_{k=1}^{N}
M_{q_k,q_{\alpha(k)}}(\mathbf{x}_i,\mathbf{x}_j),
\label{eq:relative_perm_overlap}
\end{equation}
with the phase-dependent weight function defined as
\begin{equation}
C(\alpha,\phi)
=
\sum_{\sigma \in S_N}
e^{i\phi[\pi(\sigma)-\pi(\alpha \circ \sigma)]}.
\label{eq:C_alpha_phi}
\end{equation}
Finally, the generalized particle quantum kernel is defined as the squared fidelity between the encoded states, formulated as
\begin{equation}
K_{\phi}(\mathbf{x}_i, \mathbf{x}_j)
=
\left| \langle \Phi_{\phi}(\mathbf{x}_i) | \Phi_{\phi}(\mathbf{x}_j) \rangle \right|^2 .
\label{eq:frac_kernel}
\end{equation}

Evaluating these entries produces the Gram matrix $\mathbf K_{\phi}$, which serves as a precomputed kernel for a classical support-vector machine~\cite{boser1992svm,cortes1995support}.  
For a multiclass classification task with labels $y_i \in \{1,\ldots,L\}$, we use a one-vs-rest multiclass SVM with a precomputed kernel.
The classifier assigns a decision score $f_c(\mathbf{x})$ to each class $c$, and predicts
\begin{equation}
\hat y(\mathbf{x})
=
\arg\max_{c \in \{1,\ldots,L\}} f_c(\mathbf{x}) .
\end{equation}

\section{Effective Dimension for Quantum Kernel}

The representational capacity of a quantum kernel depends on the Hilbert-space subspace explored by its encoded states.
This accessible space sets a basic, data-independent limit on what the kernel can represent before any particular learning task is specified. In multiparticle photonic systems, this space is not arbitrary. It is constrained by exchange statistics~\cite{messiah1964symmetrization}.
For $N$ distinguishable particles in $M$ modes, the full tensor-product space has dimension $M^N$. Identical particles explore much smaller symmetry sectors. Bosons are confined to the symmetric subspace, whose dimension is $\binom{M+N-1}{N}$, whereas fermions are confined to the antisymmetric subspace, whose dimension is $\binom{M}{N}$. The fractional-statistics feature map relaxes this binary constraint. By introducing a fractional exchange phase, it allows the encoded states to move beyond the purely bosonic or fermionic sectors. To isolate the role of exchange statistics, we use a statistics-only diagnostic based on Haar twirling over the unitary group $U(M)$~\cite{haartwirling}. To remove dependence on the data distribution and encoding scheme, we replace $U(\mathbf{x})$ with a Haar-random unitary $U$. For a given exchange phase $\phi$, the density matrix averaged by the ensemble is defined as
\begin{equation}
    \rho(\phi)
    =\mathbb{E}_{U\sim \mathrm{Haar}}\left[|\Phi_{\phi}(U)\rangle\langle \Phi_{\phi}(U)|\right].
    \label{eq:haar_rho}
\end{equation}
Diagonalizing $\rho(\phi)$ gives the spectrum ${p_i(\phi)}$. From this spectrum we compute the von Neumann entropy
\begin{equation}
S(\phi) = -\sum_i p_i(\phi)\ln p_i(\phi).
\end{equation}
We then define the effective dimension as
\begin{equation}
D_{\mathrm{eff}}(\phi) = \exp[S(\phi)].
    \label{eq:eff_dim}
\end{equation}
This quantity estimates how many orthogonal directions are effectively accessed by the fractional-statistics feature map.

\begin{figure*}[htp]
    \centering
    \includegraphics[width=\linewidth]{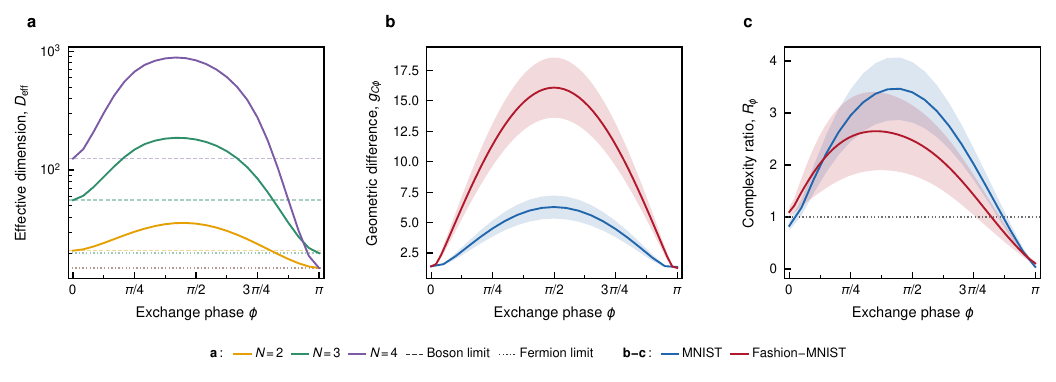}
    \caption{
    \textbf{a}, Effective dimension $D_{\rm eff}$ of the fractional-statistics feature map as a function of the exchange phase $\phi$ for different particle numbers $N$. 
    The bosonic and fermionic limits correspond to $\phi=0$ and $\phi=\pi$, respectively, whereas intermediate exchange phases access additional symmetry sectors and enlarge the effective feature space. 
    \textbf{b}, Label-independent geometric difference $g_{C\phi}$ between the fractional-statistics kernel $K_\phi$ and the distinguishable-photon baseline $K_C$. 
    This quantity characterizes the maximum geometry-level separation from the classical baseline and provides an upper bound on the model-complexity reduction that $K_\phi$ can support. 
    \textbf{c}, Realized label-dependent model-complexity ratio $R_\phi=s_C(\mathbf y)/s_\phi(\mathbf y)$ for MNIST and Fashion-MNIST. 
    Values $R_\phi>1$ indicate that the fractional-statistics kernel reduces the model complexity relative to the distinguishable-photon baseline for the given labels. 
    In \textbf{b} and \textbf{c}, solid curves denote the mean over ten independent runs, and shaded regions indicate one standard deviation.
    }
    \label{fig:main1}
\end{figure*}

The effective dimension for $N=2,3$, and $4$ particles distributed across $M=6$ modes is illustrated in Fig.~\ref{fig:main1}a. The expected boson and fermion limits are faithfully reproduced at the boundaries. In the fractional statistical region, a pronounced increase in $D_{\mathrm{eff}}(\phi)$ is observed. This enhancement becomes more pronounced for larger particle numbers, indicating that fractional phases provide access to regions of the feature space that pure bosons and fermions cannot reach.

The physical origin of this effective-dimension enhancement can be understood from the permutation sector structure of the feature state. At the bosonic and fermionic endpoints, the exchange phase reduces to the trivial and sign characters of $S_N$. For intermediate exchange phases, the phase factor is no longer a single character of $S_N$, and the feature state develops support on mixed-symmetry sectors. These sectors obey partial-Pauli occupation constraints: a sector labeled by the Young diagram $\lambda$ permits at most $n_{\max}=\lambda_1$ particles in one mode~\cite{partiialpauli}. Fractional exchange phases therefore access several sector-resolved occupation structures, opening additional feature-space directions. The corresponding central projector decomposition, sector-resolved entropy contributions, and partial-Pauli bounds are given in the Supplementary Information section III.

The effective dimension thus provides a data-independent signature of the enlarged representational capacity generated by fractional statistics. In the following, we show that a distinct kernel geometry accompanies this enlarged support and, in turn, improves classification performance.

\section{Kernel Geometry Induced by Exchange Statistics}

The effective-dimension analysis shows that fractional exchange phases enlarge the accessible feature space by relaxing symmetry constraints. A larger feature space, however, does not by itself determine whether a kernel is useful for learning. What also matters is how data points are arranged within that space. We therefore examine the geometry encoded in the Gram matrix of the fractional-statistics kernel.

To isolate the role of exchange statistics in shaping this geometry, we establish a baseline using the distinguishable-photon kernel introduced in Ref.~\cite{YinbosonKernel}. In the absence of quantum interference, this reference kernel reduces to
\begin{equation}
    K_C(\mathbf{x}_i,\mathbf{x}_j) = \sum_{\sigma\in S_N}\prod_{k=1}^{N}\left|M_{q_k,q_{\sigma(k)}}(\mathbf{x}_i,\mathbf{x}_j)\right|^2.
    \label{eq:distinguishable_kernel_collision_free}
\end{equation}

We first compare $K_\phi$ with this baseline at the level of kernel geometry alone. 
For Gram matrices, the label-independent geometric difference is defined as
\begin{equation}
    g_{C\phi} = \sqrt{\left\| K_{\phi}^{1/2}K_C^{-1}K_{\phi}^{1/2} \right\|_{\infty}
    } .
    \label{eq:geom_difference}
\end{equation}
This quantity characterizes the largest possible separation between the two kernels in the model-complexity framework of Ref.~\cite{Huangpower}. 
In this framework, the complexity of representing a label vector $\mathbf y$ with a kernel $K$ is
\begin{equation}
    s_K(\mathbf y)=\mathbf y^T K^{-1}\mathbf y .
    \label{eq:model_complexity}
\end{equation}
A smaller value of $s_K(\mathbf y)$ means that the labels are represented more simply in the feature space induced by $K$. The geometric difference bounds the corresponding complexity reduction through
\begin{equation}
    s_C(\mathbf y) \leq g_{C\phi}^{2} s_{\phi}(\mathbf y),
    \label{eq:model_complexity_bound}
\end{equation}
and therefore gives a label-independent upper bound on the improvement that the fractional-statistics kernel can provide relative to the distinguishable-photon baseline.

To determine how much of this geometric potential is realized for a given supervised task, we further compute the model-complexity ratio
\begin{equation}
    R_{\phi}=\frac{s_C(\mathbf y)}{s_{\phi}(\mathbf y)} .
    \label{eq:model_complexity_ratio}
\end{equation}
Here $R_{\phi}>1$ indicates that the fractional-statistics kernel represents the target labels with lower complexity than the distinguishable-photon kernel. The bound above implies $R_{\phi}\leq g_{C\phi}^2$.

Having established the geometric and model-complexity measures defined above, we now benchmark these quantities on two standard real-data image classification datasets, MNIST and Fashion-MNIST. A non-monotonic dependence on the exchange phase is observed in Fig.~\ref{fig:main1}b,c, indicating that the fractional-statistics regime cannot be reduced to a simple interpolation between the bosonic and fermionic endpoints. Starting from the bosonic limit, the geometric difference $g_{C\phi}$ rises rapidly as soon as a fractional exchange phase is introduced, showing that even a small departure from bosonic symmetry can substantially reshape the induced feature-space geometry. 
The maximum is reached at intermediate phases, where the fractional-statistics kernel is most strongly separated from the distinguishable-photon baseline.

The model-complexity ratio $R_\phi$ follows the same qualitative trend, with its largest values also appearing in the intermediate-phase regime. This agreement is important because $g_{C\phi}$ is independent of the labels, whereas $R_\phi$ depends on the specific classification task. Their simultaneous enhancement shows that the geometric separation induced by fractional statistics is not only a formal property of the kernel. Part of this separation is converted into a lower-complexity representation of the target labels.

\begin{figure*}[htp]
    \centering
    \includegraphics[width=\linewidth]{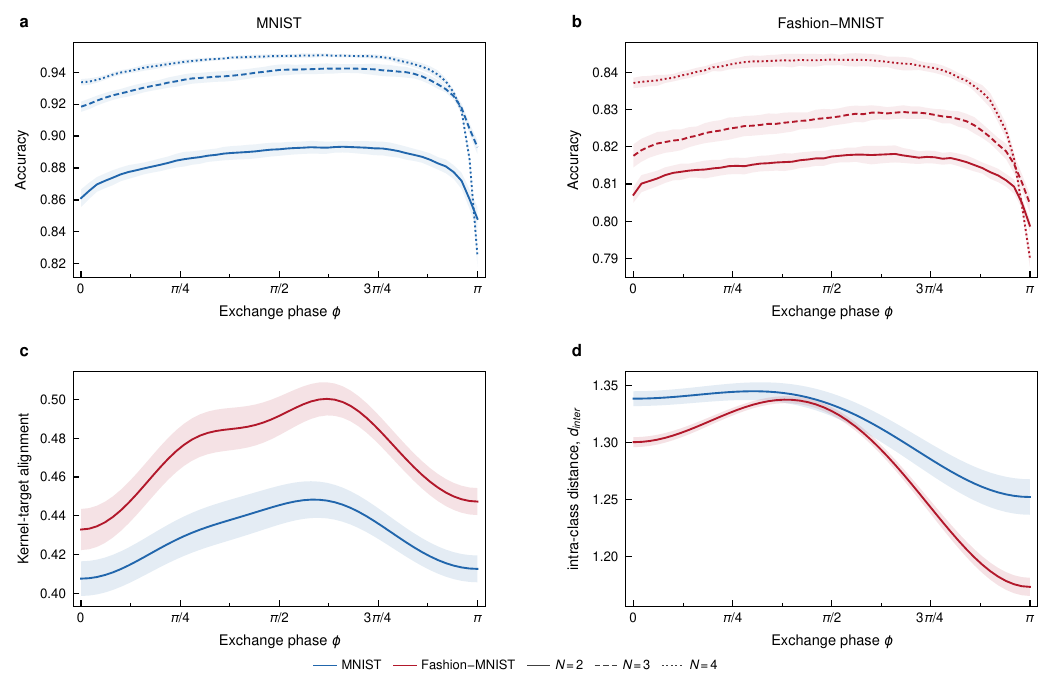}
    \caption{
\textbf{a}, Test accuracy of kernel SVM classifiers on MNIST as a function of the exchange phase $\phi$. Solid, dashed, and dotted curves correspond to $N=2,3$ and $4$ particles, respectively.
\textbf{b}, Test accuracy on Fashion-MNIST under the same kernel-SVM protocol and particle-number settings.
\textbf{c}, Kernel-target alignment between the fractional-statistics kernel $K_\phi$ and the multiclass target matrix, shown for the two-particle kernel.
\textbf{d}, Mean inter-class distance $d_{\rm inter}$ in the kernel-induced feature space for the same setting. Blue and red curves denote MNIST and Fashion-MNIST, respectively. Shaded regions indicate one standard deviation over independent data splits.}

    \label{fig:main2}
\end{figure*}

\section{Classification with Fractional-statistics Kernel}

Having characterized how exchange statistics enlarge the effective feature space and reshape the induced kernel geometry, here we ask whether these changes improve the performance of quantum kernels on classification tasks. As shown in Fig.~\ref{fig:main2}a,b, the classification accuracies on both the MNIST and the Fashion-MNIST datasets are plotted for $N=2,3$ and $4$ particles, with exchange phase $\phi \in [0,\pi]$. For both datasets, it turns out that the highest accuracies are obtained at intermediate fractional exchange phases.

In addition, Fig.~\ref{fig:main2}a,b also show that increasing the particle number improves the performance of both the bosonic and fractional-statistics kernels. This trend indicates that many-particle exchange interference can generate increasingly useful feature representations. However, the fermionic kernel behaves differently. Its performance does not improve with particle number, and the lowest accuracy is found for $N=4$ fermions. This is consistent with the effective-dimension analysis: the fermionic limit remains strongly constrained by antisymmetry and therefore accesses a smaller part of the available feature space. The optimal performance at intermediate exchange phases points to a favourable balance between exchange-induced interference and feature-space expressivity.

Meanwhile, we observe that the optimal phase is shifted towards the fermionic side of the fractional-statistics regime, rather than coinciding with the maxima of the geometry-only diagnostics discussed above. To understand this shift, we quantify both label alignment and class separation in the kernel-induced feature space. For a multiclass target matrix $T_{ij}=\mathbb{I}(y_i=y_j)$, where $\mathbb{I}$ is the indicator function, the kernel-target alignment (KTA) is defined through the Frobenius inner product $\langle A,B\rangle_F=\mathrm{Tr}(A^T B)$ as
\begin{equation}
\mathrm{KTA}(K_{\phi},T)
=
\frac{\sum_{i,j}(K_{\phi})_{ij}T_{ij}}
{\sqrt{\sum_{i,j}(K_{\phi})_{ij}^{2}}\sqrt{\sum_{i,j}T_{ij}^{2}}} .
\label{eq:kta_definition}
\end{equation}
We also evaluate the separation between samples using the kernel-induced distance
\begin{equation}
D_{ij}^{(\phi)}
=
\sqrt{(K_{\phi})_{ii}+(K_{\phi})_{jj}-2(K_{\phi})_{ij}} .
\label{eq:kernel_distance}
\end{equation}
The mean inter-class distance $d_{\mathrm{inter}}$ is obtained by averaging $D_{ij}^{(\phi)}$ over all sample pairs belonging to different classes, $\mathcal{P}_{\mathrm{diff}}=\{(i,j):i<j,\ y_i\neq y_j\}$. These quantities measure, respectively, how well the kernel is aligned with the target labels and how strongly different classes are separated in the induced kernel space~\cite{kta,distence}. As shown in Fig.~\ref{fig:main2}c,d, both quantities favour an intermediate exchange phase biased towards the fermionic side, matching the regime in which the highest test accuracy is observed.

Physically, the fermionic endpoint suppresses class separation through Pauli exclusion, whereas the bosonic endpoint enhances separation through bunching but yields weaker agreement with the target labels. Fractional statistics balance these two limits, preserving substantial inter-class separation through partial bunching while reshaping the kernel into a more label-compatible similarity structure through partial exclusion. The concurrent enhancement of KTA and $d_{\rm inter}$ therefore suggests that the classification advantage arises from a fractional-statistics kernel geometry that is both well separated and well aligned with the learning task. Photon-number-resolved analyses of KTA and class-distance geometry for $N=2,3$ and $4$ are provided in the Supplementary Information section IV.

\section{Discussion and Outlook}

In this work, we identify exchange statistics as a structural ingredient for quantum machine learning. 
The analysis proceeds from data-independent geometry to supervised learning performance. Using Haar-random unitary transformations, we first show that fractional exchange phases enlarge the effective feature space. For real-world datasets, this enlarged space is accompanied by a distinct kernel geometry. The fractional-statistics kernel displays the strongest label-independent separation from the distinguishable-photon baseline and can reduce the model complexity required to represent the observed labels. In classification tasks, the same intermediate fractional-statistics regime yields higher test accuracy and stronger kernel-target alignment on MNIST and Fashion-MNIST. Together, these results show that geometric diagnostics, evaluated before supervised training, can provide useful guidance for designing quantum kernels with improved predictive performance.

The fractional-statistics kernels studied here are also compatible with near-term integrated-photonic platforms. Entangled photon inputs with a tunable relative phase have already been used to emulate a continuous interpolation between bosonic and fermionic behavior under linear optical evolution~\cite{matthews2013entangle}, including on programmable silicon photonic processors~\cite{Qianggeneralwalk}. The two-particle regime is therefore directly accessible with existing technology. Extending the approach to larger particle numbers will require multipartite entangled inputs and mode-matched optical networks, but it does not require physical anyons or real-space braiding.

The encoding architecture provides another important direction for future work. Here we used a universal Clements interferometer as a generic photonic encoder. This is experimentally natural, but universality alone does not guarantee that the resulting feature map is well matched to a given learning problem. More structured encodings, including shallow, hierarchical, locality-preserving or trainable photonic circuits, may provide stronger inductive biases while retaining exchange-induced interference. More broadly, our results suggest that exchange structure should be treated as a tunable design variable in quantum machine learning, alongside system size, circuit architecture and data encoding.

\section*{Acknowledgments}
We thank Junkai Dong for helpful discussions. 
Z.Y. was supported by the Beijing Institute of Technology Research Fund Program under Grant No. 2024CX01015, National Natural Science Foundation of China under Grant No. 12441502, and the Fundamental Research Funds for the Central Universities.
W.L. was supported by the National Natural Science Foundation of China under Grant No. 12505028, and the Science Research Project of Hebei Education Department under Grant No. QN2025054. 
Z.W. was supported by National Natural Science Foundation of China (Grant Nos. 62272259 and 62332009), Beijing Natural Science Foundation (Grant No. Z220002), and Beijing Science and Technology Planning Project (Grant No. Z25110100810000).

\newpage
\appendix
\section{Photonic data encoding and numerical evaluation protocol}

\subsection*{Classical data preprocessing and photonic feature encoding}

The photonic encoder was implemented as a programmable Clements-type interferometer composed of a sequence of Mach-Zehnder interferometers (MZIs)~\cite{clements16}.
Since the original image data have a much higher dimension than the number of independently tunable parameters in the interferometer, we first applied PCA to obtain a compact feature vector $\mathbf{x}$. After PCA, each retained feature component was normalized to the interval $[0,1]$ before being assigned to the optical parameters.
For the $\ell$th MZI acting on modes $(m_\ell,n_\ell)$, two assigned feature components determine the beamsplitter angle and internal phase according to $\theta_\ell=\pi x_{2\ell-1}$ and $\varphi_\ell=2\pi x_{2\ell}$, respectively.
The global unitary $U(\mathbf{x})$ is obtained by cascading the corresponding MZI transfer matrices according to the chosen interferometer layout, with an individual MZI represented as

\begin{equation}
T_{m,n}(\theta_\ell, \varphi_\ell) = \begin{pmatrix} e^{i \varphi_\ell} \cos(\theta_\ell) & -\sin(\theta_\ell) \\ e^{i \varphi_\ell} \sin(\theta_\ell) & \cos(\theta_\ell) \end{pmatrix}.
\end{equation}

\subsection*{Numerical simulation setup and evaluation protocol}
All numerical experiments were simulated using a linear optical network comprising six spatial modes. To isolate the effects of exchange statistics in the effective-dimension analysis, the initial occupation strings were configured according to the total particle number: we utilized the input state $|0,0,1,1,0,0\rangle$ for two particles, $|0,0,1,1,1,0\rangle$ for three particles, and $|0,1,1,1,1,0\rangle$ for four particles. 

The learning tasks were evaluated under two distinct data configurations. To compute the geometric-difference and model-complexity ratio, we constructed binary classification tasks restricted to classes 0 and 3 from both the MNIST and Fashion-MNIST datasets~\cite{lecun-mnist,Fasion-MNIST}. Each of these random instances comprised 100 class-balanced samples. To assess the full supervised classification performance — specifically, accuracy, kernel-target alignment, and inter-class distance — the evaluation was expanded to include all 10 classes from both datasets. In these multiclass scenarios, the models were trained on 6,000 class-balanced samples and subsequently evaluated on 1,000 class-balanced test samples. For classification evaluations, the input occupation strings followed the same particle-number-dependent configurations used in the effective-dimension analysis.

To ensure robust statistical reliability, ten independent random dataset splits were generated for every task. All reported graphical curves display the mean values across ten independent splits, with the corresponding standard deviations to accurately quantify run-to-run variability.

\section{Scope of the optical fractional-statistics feature map}

For two photons, the optical simulation reproduces the exchange-phase-dependent interference expected for Abelian anyonic exchange at the level of measured correlation probabilities. For $N>2$, the construction should be interpreted more narrowly as an anyon-inspired feature map defined on particle permutations. Unlike physical anyons, which are described by braid-group representations, our model does not resolve braid histories, exchange orientations, or winding processes that can carry distinct topological information even when they induce the same final particle permutation. Instead, it retains only the induced element of the permutation group $S_N$ and assigns an exchange phase through the inversion number $\pi(\sigma)$.

This permutation-level formulation is a simplification of the full braid-group description of physical anyons. Nevertheless, it retains the feature of fractional statistics most relevant to the present kernel construction. By reducing the braid structure to permutations, the model focuses on the role of fractional exchange phases themselves. This simplification also makes the theoretical analysis more transparent and provides a direct route to a well-defined quantum kernel. The permutation-level description further implies the phase symmetry $K_{\phi}=K_{-\phi}$. Thus, although the model does not capture the full topological structure of physical anyons, it preserves the central exchange-statistical ingredient needed here and offers a controlled framework for studying how fractional statistics affects quantum-kernel geometry and learning performance.

\bibliography{ref.bib}

\end{document}